\documentclass[12pt]{article}

\usepackage{times}

\usepackage{units}

\usepackage{graphicx}

\usepackage{xcolor}

\usepackage{setspace}

\usepackage{amsmath}

\usepackage{amssymb}

\usepackage{mathrsfs}

\usepackage{xspace}

\usepackage{dsfont}

\usepackage{soul}

\newenvironment{sciabstract}{%
	\begin{quote} \bf}
	{\end{quote}}

\title{\vspace{-5.0cm}Autonomous robotic nanofabrication with reinforcement learning}

\author{
	Philipp Leinen$^{1,2,3,\ast}$,
	Malte Esders$^{4, \ast}$,
	Kristof T. Sch\"utt$^{4}$,\\
	Christian Wagner$^{1,2,\ast \ast}$,
	Klaus-Robert M\"uller$^{4,5,6,\ast \ast}$,
	F. Stefan Tautz$^{1,2,3}$
	\\
	\\
	\normalsize{$^{1}$ Peter Gr\"unberg Institut (PGI-3), Forschungszentrum J\"ulich, 52425 J\"ulich, Germany	}\\
	\normalsize{$^{2}$ J\"ulich Aachen Research Alliance (JARA)-Fundamentals of Future Information Technology,}\\
	\normalsize{52425 J\"ulich, Germany}\\
	\normalsize{$^{3}$ Experimentalphysik IV A, RWTH Aachen University, Otto-Blumenthal-Stra\ss{}e,}\\
	\normalsize{ 52074 Aachen, Germany}\\
	\normalsize{$^{4}$ Machine Learning Group, Technische Universit\"at Berlin, 10587 Berlin, Germany	}\\			
	\normalsize{$^{5}$ Max Planck Institute for Informatics, 66123 Saarbr\"ucken, Germany	}\\	
	\normalsize{$^{6}$ Department of Brain and Cognitive Engineering, Korea University, Seoul, South Korea }\\	
	\normalsize{$^\ast$ These authors have contributed equally to this work}\\
	\normalsize{$^\ast$$^\ast$ To whom correspondence should be addressed; E-mail:}\\
	\normalsize{ c.wagner@fz-juelich.de; klaus-robert.mueller@tu-berlin.de.
	} \\
}

\date{}

\begin{document}
	
	\baselineskip24pt
	
	\maketitle 
	
\begin{sciabstract}
	The ability to handle single molecules as effectively as macroscopic building-blocks would enable the construction of complex supramolecular structures inaccessible to self-assembly. The fundamental challenges obstructing this goal are the uncontrolled variability and poor observability of atomic-scale conformations. Here, we present a strategy to work around both obstacles, and demonstrate autonomous robotic nanofabrication by manipulating single molecules. Our approach employs reinforcement learning (RL), which finds solution strategies even in the face of large uncertainty and sparse feedback. We demonstrate the potential of our RL approach by removing molecules autonomously with a scanning probe microscope from a supramolecular structure -- an exemplary task of subtractive manufacturing at the nanoscale. Our RL agent reaches an excellent performance, enabling us to automate a task which previously had to be performed by a human. We anticipate that our work opens the way towards autonomous agents for the robotic construction of functional supramolecular structures with speed, precision and perseverance beyond our current capabilities.
\end{sciabstract}

\section*{Introduction}

The swift development of quantum technologies could be further advanced if we managed to free ourselves from the imperatives of crystal growth and self-assembly, and learned to fabricate custom-built metastable structures on atomic and molecular length scales routinely \cite{Temirov2008,Lafferentz2009,Koch2012,Wagner2014,Aono2015,Kawai2016,Esat2018}. Metastable structures, apart from being more abundant than stable ones, tend to offer attractive functionalities, because their constituent building blocks can be arranged more freely and in particular in desired functional relationships \cite{Esat2018}. 

It is well-established that single molecules can be manipulated and arranged using mechanical, optical, or magnetic actuators \cite{Neuman2008}, such as the tips of scanning probe microscopes (SPM) \cite{Rief1997,Gimzewski1999,Weiss2011,Pavlicek2017a} or optical tweezers \cite{Stigler2011,Gluckstad2017}. With all these types of actuators, a sequence of manipulation steps can be carried out in order to bring a system of molecular building blocks into a desired target state. The problem of creating custom-built structures from single molecules can therefore be cast as a challenge in robotics.

In the macroscopic world, robots are typically steered using either human expert knowledge or model-based approaches \cite{Atkeson1997, Deisenroth2015,Levine2016,Andrychowicz2017}. Both strategies are not available at the nanoscale, because on the one hand, human intuition is largely trained on concepts like inertia and gravity, which do not apply here, while on the other hand atomistic simulations are either too imprecise to be helpful or computationally too expensive to generate the large amount of sufficiently accurate data required for training. This is aggravated by the fact that actuators have variable and typically unknown structures and properties at the nanometer scale, making it extremely difficult if not impossible both to cover all relevant configurations of the actuator in the simulation and to establish a connection between the actual robotic process and the simulation. This leaves {\it autonomous} robotic nanofabrication as the preferred option.

In the current study, we show for the first time that Reinforcement Learning (RL) can be used to automate a manipulation task at the nanoscale. In RL, a software agent is placed in an environment at time $t=0$ and sequentially performs actions to alter the state $s_t$ of this environment\cite{Sutton2018,Carleo2017}. While executing random actions in the beginning, the agent will, based on its accumulated experience, incrementally learn a policy $\pi$ for choosing actions $a_t$ that maximize a sum over reward signals $r_{t+1}$. The reward signal is returned by the environment in a manner specified by the experimenter beforehand. The experimenter designs the reward signal such that behavior which solves the problem yields a high reward, whereas failure to do so yields a low reward. The advantage of this approach is that the experimenter does not have to specify how the agent needs to act, but instead only has to define the desired outcome.

\begin{figure*}[h!]
	\centering
	\includegraphics[width=15.5cm]{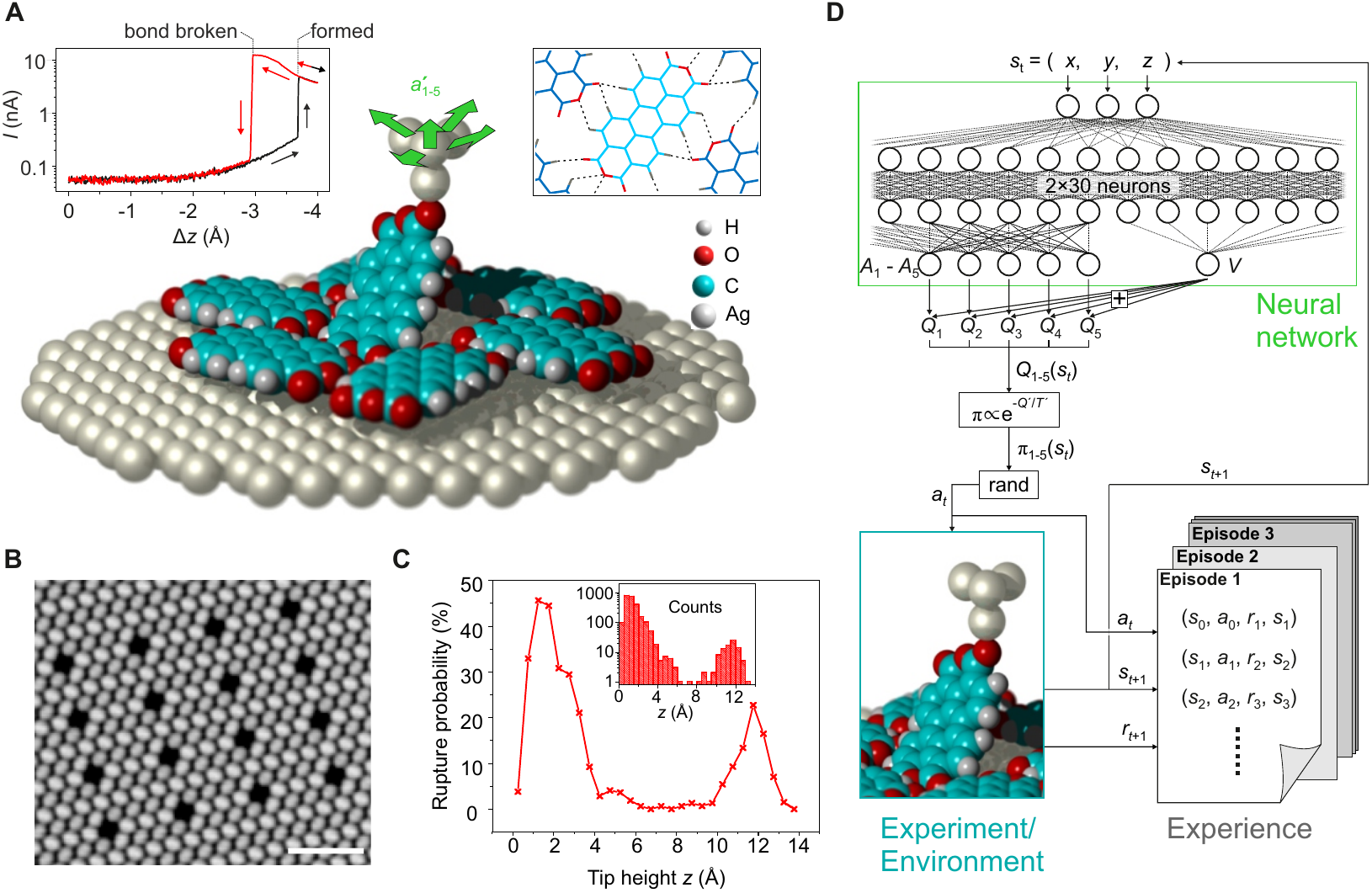}
	\sffamily
	\caption{
		\footnotesize \setstretch{1.0}\textbf{Subtractive manufacturing with a RL agent.} 
		(\textbf{A}) PTCDA molecules can spontaneously bind to the SPM tip and be removed from a monolayer upon tip retraction on a suitable trajectory. Bond formation and breaking cause strong increases or decreases in the tunneling current (left inset). The removal task is challenging, since PTCDA is retained in the layer by a network of hydrogen bonds (dotted lines in right inset). The RL agent can repeatedly choose from the five indicated actions $a'_\text{1-5}$ (green arrows) to find a suitable trajectory (action set $\mathcal{A}$: $\Delta z = \unit[0.1]{\text{\AA}}$ step plus $\unit[\pm 0.3]{\text{\AA}}$ step in $x$ or $y$ direction, or no lateral movement).
		(\textbf{B}) STM image of a PTCDA layer with 16 vacancies created by the RL agent (scale bar is $\unit[5]{nm}$).  
		(\textbf{C}) Probability of bond rupture in intervals of $\unit[0.5]{\text{\AA}}$ around tip height $z$ as a function of $z$, based on all bond-breaking events accumulated during the RL-agent experiments (inset).
		(\textbf{D}) The $Q$-function is approximated by a neural network with 30 neurons in the first, and $2\times 15$ neurons in the second hidden layer. This dueling network architecture \cite{Wang2016} features separate outputs $A_i$ and $V$, with $Q_i = V + A_i$ for actions $a'_{i=1\ldots 5}$. The actually performed action is then randomly chosen from $\mathcal{A}$ with probabilities computed with the policy $\pi$.
		\normalsize
	}
	\normalfont 
	\label{Fig1} 
\end{figure*}

Considering a compact layer of PTCDA (3,4,9,10-perylene-tetracarboxylic dianhydride) on an  Ag(111) surface, we define removing individual molecules from this layer using an SPM as the RL agent's goal (see Fig.~\ref{Fig1}). This task is an example of a subtractive manufacturing process that starts from a self-assembled molecular structure. We note that subtractive manufacturing has been identified as key to nanoscale fabrication \cite{Green2014,Kocic2019}.

\section*{Results}

\subsection*{Robotic nanofabrication as a Reinforcement Learning challenge}
A RL task is usually modeled as a Markov Decision Process (MDP) \cite{Sutton2018}, which is a Markov Process equipped with an agent which can perform certain actions at each state to influence the transition into the next state. In nanofabrication, a complete numerical representation of the environment state $\bar{s}_t$ would consist of the coordinates of all relevant atoms in the environment. The probability distribution $p(\bar{s}_{t+1},r_{t+1}|\bar{s}_t,a_t)$, which defines the probability to reach state $\bar{s}_{t+1}$ and receive reward $r_{t+1}$ after taking action $a_t$ in state $\bar{s}_t$ is deterministic at low temperatures (in our example at $T=\unit[5]{K}$), and stochastic at temperatures where thermal fluctuations are enabled \cite{Stigler2011}. The complete state $\bar{s}_t$ of the environment is generally not observable at the current level of technology \cite{Stigler2011}. To apply RL to nanofabrication we therefore suggest using an \textit{approximate} state definition $s_t$. There are several plausible elements to construct such a definition. First, there is the known actuator position. Second, there are typically \textit{some} measurable quantities (like forces) in any robotic nanofabrication setup which are functions of the complete state $\bar{s}_t$ of the environment and which could thus be used to approximate this state.

Importantly, any approximate state definition will be of much lower dimensionality than the complete state definition, such that transitions $\bar{s}_t \rightarrow \bar{s}_{t+1}$ in the complete state space cannot be captured fully by transitions $s_t \rightarrow s_{t+1}$ in the approximate state space. Hence, two states $s_t$ and $s_t'$ could be identical even when the underlying complete states $\bar{s}_t$ and $\bar{s}_t'$ are not. Using an approximate state description has several consequences: Firstly, it could break the Markov property, because the history $s_0, \ldots, s_t$ could provide more information about $\bar{s}_t$ than $s_t$ alone. Secondly, the problem could become effectively non-stationary because a change in the actuator (i.e., in the arrangement of its atoms) could change the entire probability distribution $p(s_{t+1},r_{t+1}|s_t,a_t)$, without the approximate state definition capable of capturing these changes. This could render the accumulated experience at least partially worthless. An additional source of non-stationarity of nanofabrication systems are parameters of the (external) macroscopic environment (the apparatus, the room, etc.), which are also varying slowly.  

This non-stationarity is at the heart of the difficulty of autonomous nanofabrication, because it means that the successful policy is not static but must be evolved constantly. Further, the speed at which a policy is learned needs to be faster than the rate at which $p(s_{t+1},r_{t+1}|s_t,a_t)$ changes. In practice, this requires a substantial speed-up of the standard RL algorithm, which is typically very slow because it needs a lot of training data. If this key advance was achieved, a policy $\pi(s_t)$ could be learned in the lifetime of the distribution $p(s_{t+1},r_{t+1}|s_t,a_t)$. Moreover, the intrinsically adaptive RL agent would be able to deal with occasional hidden changes of $\bar{s_t}$. Below, we demonstrate that RL can be sped up sufficiently to solve our exemplary nanofabrication task.

\subsection*{The PTCDA lifting task as a Reinforcement Learning challenge}
We have previously studied the removal of single PTCDA molecules from condensed layers on Ag(111) by manual tip control\cite{Green2014,Leinen2015,Wagner2015,Leinen2016,Temirov2018,Wagner2019}. Using motion capture and virtual reality, we were able to identify specific three-dimensional tip trajectories that reach the target state in which the molecule is fully disconnected from the surface but still bonded to the tip. We stress that the bond between one of the carboxylic oxygen atoms and the tip (Fig.~\ref{Fig1}A) typically ruptures if a random retraction trajectory is chosen, since along most trajectories the molecule-surface and intermolecular forces together exceed the strength of the tip-molecule bond. Thus, to be successful, the RL agent has to find trajectories on which the retaining force, which holds the molecule in the layer, never surpasses the tip-molecule bond strength.

In our example, neither the atomic coordinates of the object system (PTCDA layer and manipulated PTCDA molecule) nor the atomic structure of the rest of the environment (SPM tip apex) are known precisely. For the definition of the approximate state $s_t$, we could exploit two measurable quantities, the tunneling current and the force gradient of the SPM. We tried to use these quantities together with the Cartesian coordinates of the tip apex as the state description of the environment (5 dimensions), but we had to conclude that, given the limited time until the task needs to be solved, the measured quantities have too high variance to be helpful in our RL setup (see Methods). We therefore chose to significantly reduce the state description and only include the Cartesian coordinates of the tip (3 dimensions) in the state definition. Since the manipulated PTCDA molecule could assume different conformations at identical SPM tip positions if different tip trajectories led to this position, this approximate state definition is not guaranteed to have the Markov property. Because the given nanofabrication task could nevertheless be solved successfully, we chose to keep the complexity in our proof-of-concept study as low as possible and did not use strategies to attempt to restore the Markov property (see e.g. \cite{Sutton2018} ch. 17).

In the PTCDA lifting task, the non-stationarity of $p(s_{t+1},r_{t+1}|s_t,a_t)$ discussed above arises for example from abrupt changes in the atomic structure of the tip apex and from a slow drift of the piezo tubes controlling the SPM tip. Such changes influence both the measurable quantities as well as the trajectories on which it is possible to lift the molecule and thus also the rewards $r_{t+1}$ in the distribution $p(s_{t+1},r_{t+1}|s_t,a_t)$. 

It should be noted that despite the non-observability of the complete state, two important key events can be unambiguously detected: the undesired loss of contact to the tip and the desired loss of contact to the surface. In the former event, the tunneling current drops by at least an order of magnitude (Fig.~\ref{Fig1}A) which is well outside the range in which the current varies while the bond is still in place (see \cite{Fournier2011} for the case of an isolated molecule). Additionally, in both events, a clear signal can be observed in the SPM force gradient channel.

\subsection*{Formal Reinforcement Learning setup}

\begin{figure*}[h!]
	\centering
	\includegraphics[width=15.5cm]{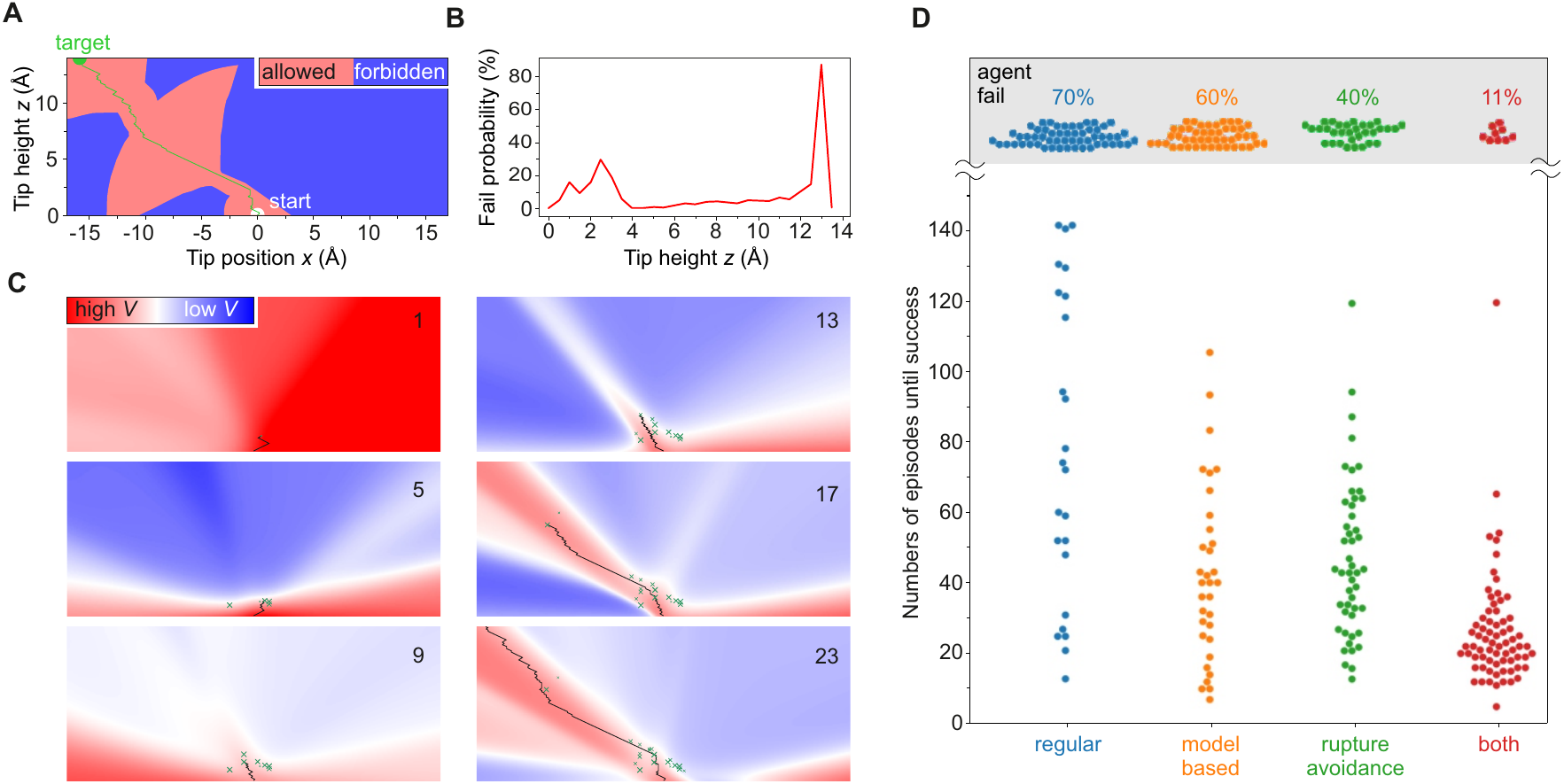}
	\sffamily
	\caption{
		\footnotesize \setstretch{1.0}\textbf{ 
			Training and performance of RL agents.
		} 
		(\textbf{A}) Map (2D slice through 3D system) of the synthetic bond rupture criteria used to study the RL agent's behavior under controlled conditions. The criteria are based on a successful experimental trajectory around which a corridor of variable diameter has been created (light red) beyond which the bond ruptures (blue). The corridor diameter is chosen to approximately reproduce the experimental bond-rupture probabilities (Fig.~\ref{Fig1}C). One successful trajectory (see panel C) is indicated in green. 
		(\textbf{B}) Probability of agent failure in $z$-intervals of $\unit[0.5]{\text{\AA}}$ in the simulation in panel A.
		(\textbf{C}) Learning progress of one RL agent. The six plots show 2D cuts ($y=0$) through the color-encoded value-function $V$ after the number of episodes indicated in the upper right corner. A 2D projection of the agent's trajectory in each episode is shown as a black line. Crosses indicate bond-breaking events triggered according to the criteria in panel A (see Supplementary Animation for a 3D view). (\textbf{D}) Swarm plot comparing the performance of different types of RL agents acting in the simulation (panel A). Plotted is the number of episodes $n$ required to accomplish the removal task for four sets of 80 simulated experiments each. A run was considered a failure after 150 unsuccessful episodes. The respective probabilities of agent-failure are indicated in the upper part of the graph.
		\normalsize
	}
	\normalfont 
	\label{Fig2} 
\end{figure*}

The RL agent steers the SPM with its actions. The environment are the actuators of the SPM and the object system. At time step $t$, the environment is in a state which we represent numerically by $s_t \in \mathcal{S}$. As noted above, we simplify the state representation to $\mathcal{S} \subset \mathds{R}^3$, and the three components are the Cartesian coordinates of the tip apex. Based on the received state, the agent picks an action $a_t$ from the set of actions $\mathcal{A}$. We specify $\mathcal{A}$ to consist of five possible actions, all of which move the tip in different directions (see Fig.~\ref{Fig1}A). The performed action in turn causes the environment to emit a new state $s_{t+1}$ and also a reward $r_{t+1} \in \mathds{R}$. 

We design the reward system as follows: If the environment transitions to a nonterminal state, we assign a default reward of $r_{t+1} = 0.01$ (see Methods for a discussion). If transitioning into a state in which the SPM tip loses contact with the molecule, the agent is penalized with $r_{t+1}= -1$ and the current episode stops. Finally, if transitioning into a state where the molecule has been lifted successfully, we assign a reward of $r_{t+1} = +1$ and the episode also stops.  After each failed episode, the molecule, by virtue of the hydrogen bonds (Fig.~\ref{Fig1}), drops back to its original position in the PTCDA-layer and the tip is moved back to $s_0=(0,0,0)$, where the tip-molecule bond re-establishes such that the next episode can start with identical conditions (provided that no change in the tip apex has occurred). 

Central to RL is the $Q$-value function, which is learned (Fig.~\ref{Fig2}C) and which, in our case, is approximated by a neural network (NN) (Fig.~\ref{Fig1}D). $Q(s_t, a_t)$ is the agent's estimate of the expected discounted future reward $G_t = \sum_{k=t+1}^{T} \gamma^{k-t-1} r_{k}$ it will receive when performing action $a_t$ in state $s_t$ and afterwards following its policy $\pi$. In a given state, the policy $\pi$ assigns action-selection probabilities to each action depending on their respective $Q$-values. In our case, $\pi$ is computed from $Q' = -Q$ using the Boltzmann distribution

\begin{equation}
	\pi(s_t, a_t, T) = \exp{\bigg(-\frac{Q'(s_t, a_t)}{T}\bigg)} \Bigg/  \sum_{a \in \mathcal{A}} \exp{\bigg(-\frac{Q'(s_t, a)}{T}\bigg)}.
	\label{eq:policy}
\end{equation}

\noindent As is common in RL, $Q$ appears with opposite sign in this equation, because a high $Q$ means a high probability, opposite to the energy/occupation relation for which this distribution was initially derived. The ``temperature'' parameter $T$ determines how greedily the agent chooses actions having higher $Q$-values.

\begin{sloppypar}
The interaction with the environment is organized into state-action-reward-state tuples $(s_t, a_t, r_{t+1}, s_{t+1})$ and stored in an experience memory to be used for training (Fig.~\ref{Fig1}D). During training, the $Q$-values predicted by the NN are adjusted to the discounted future rewards (Fig.~\ref{Fig2}C). We use an off-policy variant (see below) of the Expected SARSA (State-Action-Reward-State-Action) \cite{VanSeijen2009} algorithm, for which the loss is computed as
\end{sloppypar}
	
\begin{equation}
	L(s_t, a_t) = \Big(Q(s_t,a_t) - \big(r_{t+1} +  \gamma \sum_{a \in \mathcal{A}} \pi(s_{t+1}, a) Q(s_{t+1},a)\big)\Big)^2
	\label{eq:loss}
\end{equation}

\noindent and used to optimize the NN weights via gradient descent with samples obtained by prioritized sampling \cite{Schaul2016} from the experience memory. Note that the discounted ($\gamma=0.97$) future reward at $t+1$ is given by the $Q$-value function itself. This recursive formulation, called temporal-difference learning, allows to learn $Q$-values particularly efficiently and propagate them through the state space \cite{Sutton2018}.

\subsection*{Simulation results and RL adjustments}

Before connecting the RL agent to the microscope, we benchmarked our RL setup on a simulated system using synthetic bond-breaking criteria (Fig.~\ref{Fig2}A) derived from prior lifting experiments \cite{Green2014}. Note that the probability of bond-rupture as a function of tip height is similar between the simulation and the real experiment (Fig.~\ref{Fig1}C and Fig.~\ref{Fig2}B). Specifically, there are two heights at which there is an increased chance of rupture in the experiment, and our synthetic bond-breaking criteria recover this pattern. Even in this stable simulation with no uncontrolled variability (and complete observability), the agent typically requires more than 150 episodes to find a successful policy (Fig.~\ref{Fig2}D). As discussed above, this low data efficiency would make it (almost) impossible to achieve the goal in the real-world experiment, where we expect a substantial degree of variability over this time scale of hundreds of episodes, rendering much of the collected experience worthless.

Driven by the need to solve the task more efficiently, we introduce two modifications to the standard Expected SARSA algorithm: First, we make use of our purely Cartesian-coordinate state description to perform model-based RL similar to the Dyna algorithm \cite{Sutton1991}. Dyna uses both actual experience from interaction with the environment, and experience obtained from a learned environment model to update the $Q$-values. In our case, learning an environment model is easy, because the state transition from $s_t$ given $a_t$ to $s_{t+1}$ is deterministic. The environment model also needs to model the obtained reward. We implement our learned environment model such that it emits the default reward $r_{t+1} = 0.01$ unless the successor state $s_{t+1}$ is a known failure state (bond previously ruptured at this position in the experiment), in which case it emits the failure-reward $r_{t+1} = -1.0$. We use this environment model to sample $(s_t, a_t, r_{t+1}, s_{t+1})$ tuples around states obtained from prior experience (see Methods) and train our neural network with them. Second, we introduce a rupture avoidance mechanism by setting a negative temperature $T_{\text{train}} < 0$ in Eq.~\ref{eq:policy} during training. Using a negative temperature gives {\it lower} $Q$-values at time step $t+1$ in Eq. \ref{eq:loss} {\it more} importance. Therefore, information about impending failure states is propagated much further towards previous positions and the agent can use this information to avoid them. Of course, while acting in the environment we still set a positive temperature $T_{\text{act}} > 0$.

We next benchmark the performance of the RL setup with our two modifications in the simulation. Fig.~\ref{Fig2}D shows that especially in combination, the two modifications speed up the learning process dramatically, to the extent that it now becomes possible to connect this modified RL agent to the real-world experiment.

\subsection*{SPM setup}
While the RL agent controls our ultra-high-vacuum low-temperature non-contact AFM/STM fitted with a qPlus tuning-fork sensor \cite{Giessibl2019}, the tunneling current through the junction ($V_\text{bias} = \unit[10]{mV}$) is continuously monitored by our software (see Methods). When the bond to the tip ruptures, the increased tunneling barrier leads to a sudden drop in current and the failure-reward is assigned. Since the length of the molecule is known ($\unit[11.5]{\text{\AA}}$), the target state can also be automatically detected as any state with $z>\unit[14]{\text{\AA}}$ and the contact to the molecule still in place. Thus, the agent works autonomously until the point where the final success of the manipulation as reported by the agent is verified by the experimenter, who deposits the molecule from the tip elsewhere onto the Ag(111) surface \cite{Green2014} and images the vacancy that is created in the PTCDA layer (Fig.~\ref{Fig1}B). Tip changes which occasionally occurred either during an episode or when depositing the molecule back onto the surface have been identified by checking changes in the STM image contrast or position and in the tip-molecule bonding behavior.

\section*{Discussion}

\subsection*{Analysis of the learning process}

\begin{figure*}[h!]
	\centering
	\includegraphics[width=15.5cm]{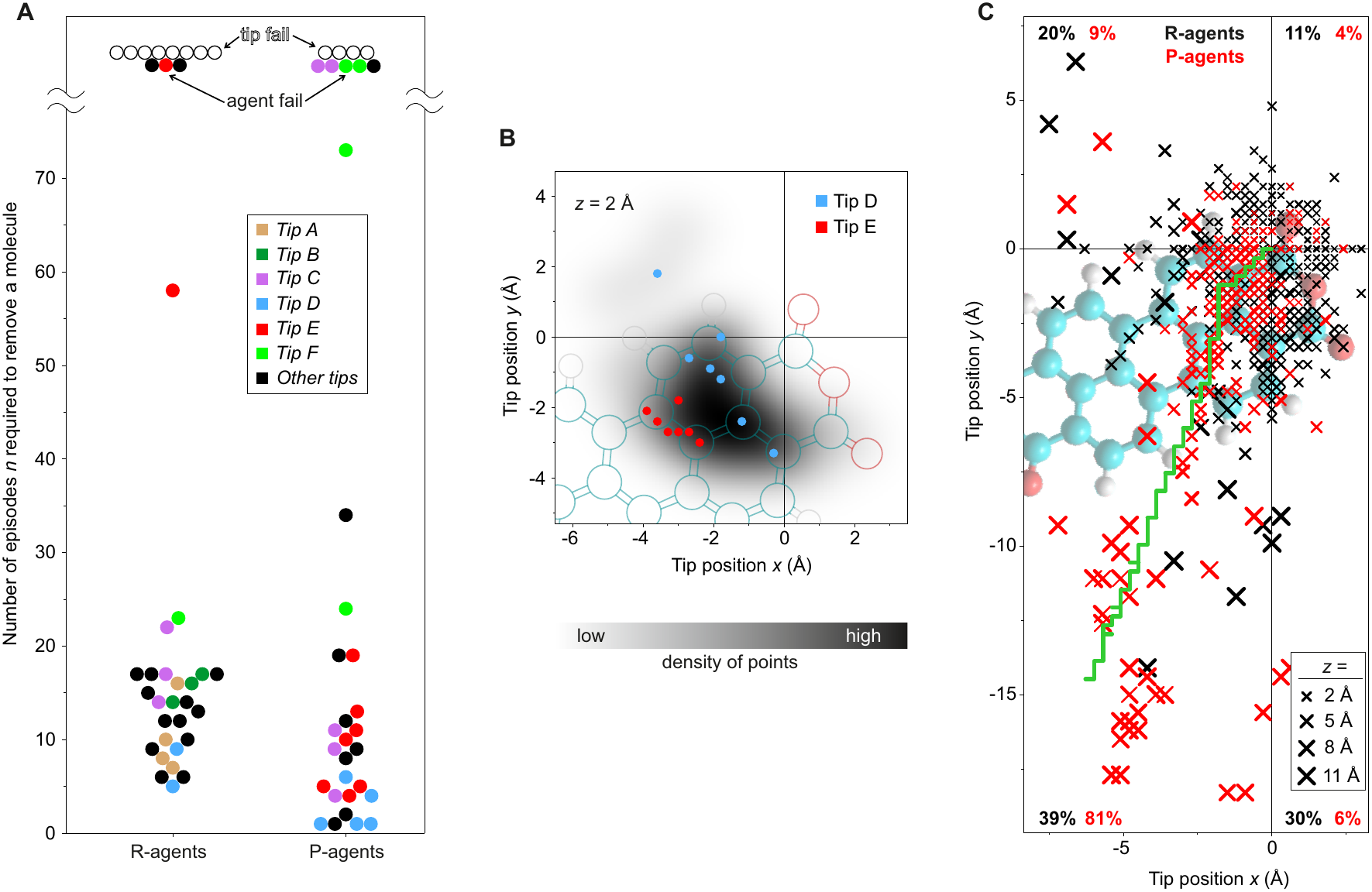}
	\sffamily
	\caption{
		\footnotesize \setstretch{1.0}\textbf{Performance of the RL agent in experiment.} (\textbf{A}) Swarm plot of the number of episodes $n$ required to accomplish the removal task. Groups of at least 3 data points acquired with the same tip are identically colored (except black). If a tip capable of removal (proven by a successful experiment) failed in another experiment, the respective data point is labeled as "agent fail". Points labeled as "tip fail" denote tips with which the removal task has never been accomplished, notwithstanding that this could, in principle, also be a failure of the agent. 
		(\textbf{B}) Density of $(x,y)$-positions where all (ultimately successful) tip trajectories pass through the $z$-region of highest bond-rupture probability ($z=\unit[2]{\text{\AA}}$, Fig.~\ref{Fig1}C). The positions for \textit{Tip~D} (strong tip) and \textit{Tip~E} (weak tip) are indicated by dots. (\textbf{C}) $(x,y)$-projections of all bond ruptures occurring within the first ten episodes for R- and P-agents. Cross sizes indicate rupture heights $z$. The quoted numbers give the percentage of rupture points located in each of the four quadrants of the coordinate system. The green curve shows the last trajectory chosen by the P-agent during its pre-training. Its direction indicates why the P-agents have a clear preference to explore the promising (Panel B) lower left quadrant, which explains their performance edge (Panel A).
		\normalsize
	}
	\normalfont 
	\label{Fig3} 
\end{figure*}

We measure the performance of each RL agent by the number $n$ of episodes which it requires to solve the removal task. To separate intrinsic RL stochasticity from the uncontrollable variability of the SPM manipulation, we plot the data points $n$ of real-world experiments that were conducted with identical tips in the same color in Fig.~\ref{Fig3}A. The scatter is indeed smaller within groups of experiments with identical tips. Moreover, the difficulty of the removal task clearly depends on the tip. For example, with tip D removal is easy, resulting on average in small $n$, while tip E, with which removal appears more difficult, results in larger $n$.

A small force threshold of the tip-molecule bond reduces the fraction of successful trajectories. For particularly weak tips (labeled \textit{tip failure} in Fig.~\ref{Fig3}A), the removal task cannot be solved at all. One would expect that for larger $n$, i.e. weaker tips, successful trajectories cluster more narrowly in space. Fig.~\ref{Fig3}B, in which we compare the $(x,y)$ coordinates of successful trajectories at $z=\unit[2]{\text{\AA}}$ for tip D and tip E, clearly confirms this effect. The distribution of the corresponding $(x,y)$ coordinates for all tips (plotted as a gray-scale background) is rather broad and indeed similar to the distribution for the strong tip D. For the weak tip E, however, the agent has to traverse a very specific region in the $xy$ plane to avoid bond-breaking. This naturally explains why tip E tends to require larger numbers $n$ of episodes until success. 

In Fig.~\ref{Fig3}A we also compare randomly initialized  (R) agents with pre-trained (P) ones. R-agents start with random weights in the neural network, while P-agents have already solved the removal task once with one particular tip. Initially, all P-agents are identical, i.e. they have the same experience and NN weights. On average, P-agents perform better than R-agents. This is evident both in the complete dataset and for individual tips, see for instance tip E. It clearly demonstrates that at least some knowledge about the removal task is universal and can be transferred to new tip configurations.

 Fig.~\ref{Fig3}C reveals the nature of this universal knowledge. In this figure, we have plotted the $(x,y)$ coordinates of the rupture points (i.e., termini of unsuccessful episodes) for R-agents (black) and P-agents (red). The data is limited to the first ten episodes of each experiment, in which the difference in training and experience between both types of agents is most significant. The plot shows that the randomly initialized agents explore all $(x,y)$ directions rather uniformly, while the pre-trained agents have a strong bias towards the lower left quadrant $(x<0,y<0)$ through which almost all successful trajectories pass (Fig.~\ref{Fig3}B). This bias is the essence of the universally valid policy which, once learned, gives P-agents a performance edge over R-agents. We note that this policy is consistent with our general understanding that molecules have to be ``peeled'' out of the condensed layer along their long axis to break the hydrogen bonds sequentially and limit the associated retaining force\cite{Green2014}.

\subsection*{Conclusion}
Automatically fabricating complex metastable structures at the molecular level is a highly desirable goal. Given the limited observability and uncontrolled variability involved, this goal seemed out of reach until now. In this proof-of-concept study, we demonstrated that indeed autonomous robotic nanofabrication becomes possible using RL without the necessity of human intervention. We chose the real-world task of lifting a molecular structure off a material surface as a textbook example. Because we used the RL framework, it was \textit{not} necessary to specify how to solve the task -- instead only the goal had to be set, which is clearly  easier. We showed that an RL agent in full control of an SPM setup is not only capable of reaching a real manipulation goal with a moderate number of trials, but that it is moreover also robust enough to transfer a previously learned policy to new object systems. Furthermore, also other nanofabrication tasks, utilizing different molecules could benefit from the accelerated learning approaches which we used to enable this performance, namely negative training temperature and model-based learning.

The limited observability is perhaps the most severe limitation of RL at the nanoscale. Although RL can indeed work under partial observability and in stochastic environments, such conditions negatively affect the number of trials needed to solve a task. To alleviate this problem, one could try using a hybrid approach in which insight from atomistic simulations is used for guiding the RL agent in its exploration of possible solutions. While atomistic conformations may not be practically accessible in detail, related measurements like tunneling current and force (gradient) are. Hence, a future research direction could focus on including such helpful variables into an RL setup. Moreover, the combination of autonomous SPM-based nanofabrication with autonomous tip preparation \cite{Rashidi2018,Gordon2019,Krull2020} would be a logical development.

In conclusion, we demonstrated that autonomous robotic nanofabrication is viable. It enables immediate progress towards the freedom of designing quantum matter, beyond the constraints of even the most complex quantum materials.

\section*{Methods}

\subsection*{Experiments}
\label{appendix:SPM}
PTCDA is deposited onto Ag(111) at room temperature and briefly annealed to $\unit[200]{^\circ C}$. The PtIr tip of the qPlus sensor was cut by ion-beam etching and prepared via indentation into the uncovered Ag(111) surface. Since each indentation typically changes the tip apex structure, it affects the strength of the molecule-tip bond and thus the difficulty of the removal task. This allowed us to test the RL agent at various levels of difficulty. The RL agent controls the tip via a voltage source the output of which is added to the piezo voltages of the SPM. 

In principle, the primary criterion to quantify the performance of an agent should be the time the agent requires to accomplish a manipulation task. In order to assess agents for simulated and real systems on equal footing, we use the number of episodes $n$ required for this task. This quantity $n$ is not fully, but closely related to the time needed (episodes may take longer or shorter depending on the length of the trajectory). Using the wall-clock time as a criterion is moreover rather meaningless, since we have intentionally slowed down the removal process in the experiment to a point where we could carefully observe the actions of the agent in order to, for example, spot changes in tip apex structure immediately. A removal process took typically 5 to 10 minutes in the experiment. The tip apex changed in 20\% of the removal experiments (not to be confused with episodes) which were excluded from the statistics. During re-deposition of the removed molecule onto the surface, the apex changed with a probability of 15\%.

\subsection*{Approximate state description}
Generally speaking, the Markov property allows a much better theoretical treatment of reinforcement learning, including, e.g., convergence proofs for the possibility of finding the optimal solution. Without the Markov property, convergence cannot be guaranteed on theoretical grounds. With our reduced state description, distinct complete states could in principle result in identical approximate states. In this case the Markov assumption would be violated, because the past trajectory could be informative about future states. The inclusion of the measured $I$ and $\Delta f$ would allow discriminating a larger set of the complete states, thus reducing the scope of hidden parameters and associated apparent memory, making the process altogether more Markovian. As a consequence, fabrication tasks which involve processes that display, e.g., a strongly hysteretic behavior would benefit from including $I$ and $\Delta f$, simply because the agent could make more informed decisions based on its ability to discriminate the respective complete states better. In principle, this could boost the learning efficiency, as the agent would not receive apparently contradicting information. But there is also a downside: Generalizing a policy across regions of similar approximate states becomes much harder if it is of higher dimensionality (e.g., 5 vs 3 dimensions), especially if $I$ and $\Delta f$ vary strongly. This hinders the learning process. Our initial RL approaches included $I$ and $\Delta f$ in the approximate state, however, for the present manipulation task, the disadvantage of increased dimensionality with strongly varying $I$ and $\Delta f$ outweighed the advantage of being able to better distinguish between underlying complete states.

\subsection*{Reinforcement Learning}
The pure state transitions $p(s_{t+1}|s_t, a_t)$ in our setup are deterministic, because the SPM tip can be moved deterministically to a new position. We use this fact to introduce model-based RL. Note that this determinism is a result of our choice of restricting the state description to the coordinates of the tip. If we included the tunneling current or the force gradient measurements, model-based RL would not be possible anymore, unless one would learn to model these variables as well, which proved too unstable in our pilot experiments.\\
In order to prevent overly optimistic estimates of the $Q$-values \cite{Thrun1993}, we use the Double-$Q$-learning approach \cite{Hasselt2010}, which also works with function approximation \cite{Hasselt2016}. Note that instead of $Q$-learning, we use Expected SARSA \cite{VanSeijen2009}. While Expected SARSA is an on-policy algorithm, we make it off-policy by using different temperatures $T_\text{train}$, $T_\text{act}$ for the Boltzmann-distribution (Eq.~\ref{eq:policy}) during training compared to when using the network to act in the environment. In Double-$Q$-learning, two networks are used in parallel: they start out with equal weights, but in the subsequent training steps, only one network is updated (the ``live network'') while the other is held fixed for a while (the ``target network''). When computing $Q$-value estimates for time step $t+1$ in the loss function (Eq. \ref{eq:loss}), the target network is used to obtain the actual $Q$-values, while the live network is used for the probabilities of the Boltzmann-distribution in the training policy (Eq.~\ref{eq:policy}) and to compute $Q(s_t,a_t)$. Every $200$ training steps, the weights of the target network are set to the weights of the live network, such that both networks once again have equal weights.

\subsection*{Rewards}
\label{appendix:rewards}
There is a trade-off between sparse and shaped rewards. The former are only given once the agent either accomplishes its final goal or ultimately fails, while the latter also reward (or punish) intermediate steps, thus directing the agent more efficiently towards its goal. If not chosen well, shaped rewards can induce unwanted agent behavior, while sparse rewards induce no such bias. We use sparse rewards because of the poor observability of the full state (atomic coordinates) of the object system, and the resulting lack of information regarding the assessment of intermediate steps. Therefore, we give a reward of $+1.0$ for success (fully lifting the PTCDA-molecule out of the surface) and $-1.0$ for rupture of the bond between the tip and PTCDA-molecule. However, we do slightly shape the reward by giving a reward of $+0.01$ for each non-terminal step of the agent. Physically, this is motivated by the fact that each step separates the molecule $\unit[0.1]{\text{\AA}}$ further from the surface. On the RL side, this small reward makes the agent prefer exploring trajectories on which it previously advanced very far before rupture. This happens because the (discounted) propagation of rewards to previous states during training (Eq.~\ref{eq:loss}) makes states along longer trajectories have higher value and therefore more ``attractive'' to the agent.

\subsection*{Neural network architecture}
The neural network used to approximate the $Q$-function consists of the three-neuron input layer receiving $(x, y, z)$ followed by a hidden layer with 30 neurons. After this intermediate layer, the network splits up into two separate streams, as inspired by the Dueling Network Architecture \cite{Wang2016}. Both streams consist of a hidden layer with 15 neurons, and an output layer with either $1$ neuron ($V$-value stream) or $5$ neurons representing the $5$ possible actions ($A$-value stream) with $Q_i = V + A_i$. The neurons in all hidden layers have tanh-nonlinearities. The number of neurons in the hidden layers was tuned empirically in the simulation to be as low as possible while maintaining optimal performance.

\subsection*{Training details}
The neural network was trained after each episode and held fixed during episodes. The optimization method was "Adam" \cite{Kingma2015} with a constant learning rate of $10^{-3}$ and a batch size of $30$. In order to avoid performing many training steps while little experience is present, the number of training steps was increased in the first ten episodes, from 200 after the first episode to finally 2000 training steps after ten or more episodes. The discount for future rewards was $\gamma = 0.97$. The training-temperature was $T_\text{train} = -0.1$, and the action-temperature used to select actions during an episode was $T_\text{act} =  0.004$.

During training, 10\% of the samples are chosen from actual experience that was obtained during any previous episode. These samples are drawn with prioritized experience replay \cite{Schaul2016}. 90\% of the samples are obtained from the environment model. The values of all parameters given above were selected via grid searches in the simulation environment.

\subsection*{Rupture avoidance mechanism}
\label{appendix:rupture_avoidance}
Prior human trials of removing the PTCDA molecule from the surface show that if a bond rupture (failure) occurs at a given location, ruptures are likely to occur in the area around it too. Also, as stated in the main text, a viable trajectory needs to be found as fast as possible, because the SPM tip might change at any time. So the agent needs to explore the state space quickly for a promising direction. Therefore, we implement a mechanism to make the agent rapidly avoid states where the tip-molecule bond previously ruptured, and particularly also the states leading up to it.

Usually, RL algorithms like the one trained with Eq.~\ref{eq:loss} tend to ``ignore'' future negative rewards, if there is a strategy that narrowly avoids them: to compute the expected future reward at time $t+1$, they weigh the highest $Q$-values more than the lower ones. In the limit as $T \to 0$, all $Q$-values but the highest one are ignored. Because of this, information about failure-states barely propagates to any states that lie two or more steps away. In our case, this means that with a randomly initialized neural network, an agent would try to lift the molecule on similar paths each episode until it is absolutely certain that this path is not viable. This leads it to fail at nearly identical locations each time and therefore it does not explore efficiently. For general RL settings, this may be the desired behavior, but we need the agent to learn as quickly as possible to avoid the failure state and the states leading up to it. Our solution is to use a negative temperature $T_\text{train}$ during training, which has the effect of inverting the importance of the $Q$-values as computed by Eq.~\ref{eq:policy}. Now, low $Q$-values, which indicate danger of rupture, are given a high importance. This information about a rupture is therefore propagated much further.

There is a trade-off here, because the further away the agent tries to stay from known failure states, the more likely it becomes that it misses a viable trajectory. We can control this trade-off by changing the training temperature. In the simulation environment, we found that the optimal training temperature was $T_\text{train}=-0.1$.

We tested whether using RL is necessary at all, then, if all the agent does is avoid the regions of rupture states. We conducted an experiment in which the agent tries a random trajectory in the first episode, and in all following episodes chooses its actions such that it stays as far away as possible from all previously occurred ruptures. Despite significant experimentation with this approach, the agent never reached the goal state but got stuck in dead ends. RL on the other hand can identify such dead ends and plan  trajectories to avoid them.

\subsection*{Exploiting the Cartesian state description for model-based planning}
\label{appendix:dyna}
We use a slight variation of Dyna \cite{Sutton1991} for model-based planning. Dyna in general updates $Q$-values with state transitions sampled from a learned environment model. To learn an environment model, we make use of the fact that the result of performing an action from a known state deterministically results in a new state, since the state description includes only the Cartesian coordinates of the tip, and each action moves the tip by a specified amount. To obtain a state from our environment model, we first pick a random state from the unique set of actually visited states, and sample a new position around it. To sample a new position, we randomly pick between zero and four actions from our action set $\mathcal{A}$ and use them to walk to a new location. At each step, we randomly either walk into the usual positive z-direction, or instead walk in negative z-direction. The new location becomes $s_t$. Then, we sample an action $a_t$ and the resulting successor state $s_{t+1}$, which is easily computed given $s_t$ and $a_t$. At this point we only need a reward $r_{t+1}$. If the sampled successor state is a known failure state (where bond breaking was observed before), the environment model emits the failure-reward of $-1$. Otherwise it emits the default step-reward $+0.01$. In this way, we generate new $(s_t,a_t,r_{t+1},s_{t+1})$ tuples and use them for training (Eq.~\ref{eq:loss}) in the same way as with regular samples. When training the NN, we use 10\% samples from real experience, and 90\% samples from the environment model. Especially in combination with the rupture avoidance mechanism, this propagates failure information to a much larger number of states, which can then be avoided in the next episodes.

\pagebreak

\section*{Acknowledgements}

\textbf{Funding:}
The authors acknowledge support from the Institute for Pure and Applied Mathematics (IPAM) at UCLA (program "Understanding Many-Particle Systems with Machine Learning"). C.W. acknowledges funding through the European Research Council (ERC-StG 757634 "CM3"). F.S.T. acknowledges financial support by the Deutsche Forschungsgemeinschaft through SFB 1083, project A12. K.-R.M. was partly supported by the German Ministry for Education and Research (BMBF) under Grants 01IS14013A-E, 01GQ1115 and 01GQ0850; the German Research Foundation (DFG) under Grant Math+, EXC 2046/1, Project ID 390685689 and by the Institute for Information \& Communications Technology Planning \& Evaluation (IITP) grant funded by the Korea government (No. 2017-0-01779). The authors thank Taner Esat for supporting the development of the RL-SPM interface software.
\textbf{Author contributions:}
C.W., K.T.S., K.-R.M. and F.S.T. conceived and designed this research during the IPAM long program program "Understanding Many-Particle Systems with Machine Learning". M.E. and K.T.S. designed and implemented the RL agent. P.L. performed the experiments. C.W., P.L. and M.E.  analyzed the data. C.W., K.-R.M and F.S.T. developed the conceptual framework. M.E., C.W. and F.S.T. wrote the paper, with input from all authors.

\end{document}